\documentclass[twocolumn,superscriptaddress,showpacs,aps,amsmath,amssymb]{revtex4}
\usepackage{epsfig}
\usepackage{amsbsy}
\usepackage{amsmath}
\usepackage{amsfonts}
\usepackage{graphicx}
\usepackage{latexsym}

\begin{document}

\title{{\Large Exact non-Markovian cavity dynamics strongly coupled to
a reservoir}}
\author{Heng-Na Xiong}
\affiliation{Department of Physics and Cenert for Quantum
information Science, National Cheng Kung University, Tainan 70101,
Taiwan}
\affiliation{Zhejiang Institute of Modern Physics and
Department of Physics, Zhejiang University, Hangzhou, People's
Republic of China}
\author{Wei-Min Zhang}
\email{wzhang@mail.ncku.edu.tw} \affiliation{Department of Physics
and Cenert for Quantum information Science, National Cheng Kung
University, Tainan 70101, Taiwan}
\author{Xiaoguang Wang}
\affiliation{Zhejiang Institute of Modern Physics and Department of
Physics, Zhejiang University, Hangzhou, People's Republic of China}
\author{Meng-Hsiu Wu}
\affiliation{Department of Physics
and Cenert for Quantum information Science, National Cheng Kung
University, Tainan 70101, Taiwan}

\date{May 6, 2010}

\begin{abstract}
The exact non-Markovian dynamics of a microcavity strongly coupled
to a general reservoir at arbitrary temperature is studied. With the
exact master equation for the reduced density operator of the cavity
system, we analytically solve the time evolution of the cavity state
and the associated physical observables. We show that the
non-Markovian dynamics is completely determined by the propagating
(retarded) and correlation Green functions. Compare the
non-Markovian behavior at finite temperature with those at
zero-temperature limit or Born-Markov limit, we find that the
non-Markovian memory effect can dramatically change the coherent and
thermal dynamics of the cavity. We also numerically study the
dissipation dynamics of the cavity through the mean mode amplitude
decay and the average photon number decay in the microwave regime.
It is shown that the strong coupling between the cavity and the
reservoir results in a long-time dissipationless evolution to the
cavity field amplitude, and its noise dynamics undergoes a critical
transition from the weak to strong coupling due to the non-Markovian
memory effect.
\end{abstract}


 \maketitle

\section{Introduction}

The dissipation quantum dynamics of optical cavities has been well
investigated and deeply understood under the Born-Markov (BM)
approximation \cite{Car93}. The BM approximation is valid when the
coupling between the system and the environment is weak enough so
that the perturbation is applied, and meantime the characteristic
time of the environment is sufficiently shorter than that of the
system so that the non-Markovian memory effect is negligible.
However, in many situations in the recent development of optical
microcavities, the strong coupling effect or the long-time memory
effect has become an important factor in controlling cavity
dynamics. Typical examples include optical fields propagating in
cavity arrays or in an optical fiber \cite{optical field in cavity
arrays,optical field in fiber -1,optical field in fiber -2}, trapped
ions subjected to artificial colored noise \cite{trapped ion in
engineered reservior -1,trapped ion in engineered reservior
-2,trapped ion in engineered reservior -3,trapped ion in engineered
reservior -4}, and microcavities interacting with a coupled
resonator optical waveguide (CROW) or photonic crystals \cite{cavity
with CROW -1,cavity with CROW -2,cavity with CROW -3,cavity with
CROW -4,cavity with CROW -5,cavity with CROW -6,cavity with CROW
-7,cavity with CROW -8}, etc. Specifically, for the trapped ions
coupled with an engineered reservoir, the change of the
characteristic frequency of the reservoir can be accomplished simply
by applying a random electric field through a band-pass filter
defining the frequency spectrum of the reservoir \cite{trapped ion
in engineered reservior -1}. While, for a cavity interacting with
CROW or photonic crystals, the coupling between them is controllable
by changing the geometrical parameters of the defect cavity and the
distance between the cavity and the CROW \cite{cavity with CROW -7}.
Both of them provide non-Markovian dissipation and decoherence
channels \cite{trapped ion in engineered reservior -2,trapped ion in
engineered reservior -3,trapped ion in engineered reservior -4,
cavity with CROW -8}. These strong coupling or long-time memory
effects result in a complicated non-Markovian process in cavity
systems that has become a crucial concern for the rapid development
of quantum information and quantum computation in terms of photons
\cite{Nie00}. The non-Markovian behavior of the trapped ions has
been discussed in many works \cite{trapped ion in engineered
reservior -2,trapped ion in engineered reservior -3,trapped ion in
engineered reservior -4}. In this paper, we shall investigate the
non-Markovian dynamics for the second case, the cavity strongly
coupled with its environment (CROW or photonic crystals).

Quantum dynamics of cavity systems is completely described by the
master equation of the reduced density operator by taking the cavity
as an open system. The master equation under the BM approximation
can be found in many textbooks \cite{Bre07,Wei99,Car93}. However, an
exact master equation beyond the BM approximation is only derived in
a few works. The first exact master equation, also called as
Hu-Paz-Zhang master equation for quantum Brownian motion, was found
almost two decades ago \cite{Hu922843}, in terms of Wigner
distribution function in phase space via Feynman-Vernon influence
functional \cite{Fey63118}. Recently, our group developed exact
master equations of the reduced density operator for both the
fermion systems \cite{Tu08235311,Tu098631} and the bosonic systems
\cite{An07042127,An090317}, by using an extended Feynman-Vernon
influence functional. These exact master equations can fully depict
the quantum dissipative and decoherence dynamics in various open
systems in the strong coupling regime. For the cavity system, the
master equation obtained in \cite{An07042127,An090317} is for
zero-temperature. Here we shall extend it to a finite temperature.
In practical, the temperature effect is also unavoidable and
non-negligible. It has been pointed out \cite{Car93} that for cavity
frequency lies in the microwave regime, thermal photons are
presented even at liquid helium temperatures \cite{experimental
parameters}.
In this paper, we shall use the exact master equation which is valid
at arbitrary temperature to investigate the exact non-Markovian
dynamics of a general cavity strongly coupled with its reservoir,
and to find general features of the coupling and temperature
dependence in non-Markovian dynamics.

The paper is organized as follows. In Sec. II, we introduce the
exact master equation  we developed recently for the reduced density
operator of the cavity system coupled to a general reservoir, from
which the second-order perturbation master equation and the BM maser
equation are reproduced at well-defined limits. In Sec.~III, we then
solve analytically the exact master equation via the coherent-state
representation, included the exact solutions of mean field amplitude
and the average photon number inside the cavity that experimentally
measurable. We also obtain explicitly the reduced density matrix for
three different initial states: the vacuum state, the coherent state
and the mixed state, to show the different non-Markovian behaviors.
The decoherence dynamics of cavity field is explicitly analyzed. In
Sec.~IV, we numerically demonstrate the exact non-Markovian behavior
through the time-dependence of the mean mode amplitude and the
average photon number in the cavity in both the weak and strong
coupling regimes for three typical spectral densities, the Ohmic,
sub-Ohmic and super-Ohmic cases, with the cavity's frequency being
focused in the microwave regime. We find that the strong coupling
between the cavity and the reservoir results in a long-time
dissipationless evolution to the cavity field amplitude, and the
noise dynamics undergoes a critical transition from the weak to
strong coupling due to the non-Markovian memory effect. Finally a
conclusion is given in Sec. V.

\section{Exact master equation for a cavity in a general
reservoir}

We consider a cavity with a single mode coupled to a general
reservoir. The Hamiltonian of the total system is given by
\begin{equation}
H=\hbar \omega_0 a^{\dag}a + \sum_{k}\hbar \omega
_{k}b_{k}^{\dag}b_{k} + \sum_{k} \hbar V_{k}\left(
b_{k}^{\dag}a+a^{\dag}b_{k}\right), \label{ht}
\end{equation}%
in which the first term is for the single cavity mode with $a^\dag,
a$ being the creation and annihilation operators of the cavity
field, $\omega_0 $ is its frequency; the second term is the
Hamiltonian $H_R$ for a general reservoir modeled as a collection of
infinite harmonic oscillators, where $b_{k}^{\dag}$ and $b_{k}$ are
the corresponding creation and annihilation operators of the $k$th
oscillator with the frequency $\omega_{k}$. The coupling between the
cavity and the environment is described by the third term and
$V_{k}$ is the coupling strength between them, which is tunable for
the reservoir being CROW or photonic crystals \cite{cavity with CROW
-1,cavity with CROW -2,cavity with CROW -3,cavity with CROW
-4,cavity with CROW -5,cavity with CROW -6,cavity with CROW
-7,cavity with CROW -8}.

\subsection{Exact master equation}

The exact master equation for the cavity field is given in terms of
the reduced density operator which is defined from the density
operator of the total system by tracing over entirely the
environmental degrees of freedom: $\rho(t) \equiv {\rm tr}_{\rm R}
\rho_{\rm tot}(t)$, where the total density operator is governed by
the quantum Liouville equation $\rho_{\rm tot}(t)=
e^{-\frac{i}{\hbar}H(t-t_0)} \rho_{\rm tot}(t_0)
e^{\frac{i}{\hbar}H(t-t_0)}$. As usual \cite{Leg871}, assuming that
the cavity field is uncorrelated with the reservoir before the
initial time $t_0$: $\rho_{\rm tot}(t_0)=\rho(t_0) \otimes
\rho_{R}(t_0)$, and the reservoir is initially in the equilibrium
state: $\rho_R(t_0)={1\over Z}e^{-H_R/(k_{\rm B}T)}$ where $k_{B}$
is the Boltzmann constant and $T$ the reservoir's initial
temperature. Then tracing over all the environmental degrees of
freedom can be easily carried out using the Feynman-Vernon influence
functional approach \cite{Fey63118} in the framework of coherent
state path-integral representation \cite{Zhang9062867}. The
resulting master equation for the reduced density operator has a
standard form similar to the master equation for electrons in
nanostructure we developed recently \cite{Tu08235311,An090317} (with
some sign difference due to the different statistical property
between fermions and bosons):
\begin{align}
\dot{\rho}& \left( t\right) =-i\omega'_0(t) \left[ a^{\dag}a,\rho(t)
\right] \notag \\ & +\kappa \left( t\right) \big\{
2a\rho(t) a^{\dag}-a^{\dag}a\rho(t) -\rho(t) a^{\dag}a \big\}  \notag \\
&+\widetilde{\kappa}(t) \big\{ a^{\dag}\rho(t) a+a\rho(t)
a^{\dag}-a^{\dag}a\rho(t) -\rho(t) aa^{\dag} \big\} ,  \label{eme}
\end{align}%
where the time-dependent coefficient $\omega'_0 (t)$ is the
renormalized frequency of the cavity, while $\kappa(t)$ and
$\widetilde{\kappa}(t)$ describe the dissipation and noise to the
cavity field due to the coupling with the reservoir. These
coefficients are non-perturbatively determined by the following
relations:
\begin{subequations}
\label{coefs}
\begin{align}
& \omega'_0 (t) = -{\rm Im}[\dot{u}(t)u^{-1}(t)] , \label{coefs-a}\\
& \kappa \left( t\right) =-{\rm Re}[\dot{u}(t)u^{-1}(t)],  \label{coefs-b} \\
& \widetilde{\kappa}(t) =\dot{v}(t)-2 v(t){\rm Re}[\dot{u}(t)u^{-1}(t)],  \label{coefs-c}
\end{align}
\end{subequations}
and $u\left( t\right) $ and $v(t)$ satisfies the integrodifferential
equations of motion:
\begin{subequations}
\label{uv-eq}
\begin{align}
\dot{u}(\tau) +i \omega_0 u (\tau)+ & \int_{t_0}^{\tau }  d\tau' g
(\tau -\tau') u (\tau') =0,  \label{u-e} \\
\dot{v}(\tau) +i \omega_0 v (\tau)+ & \int_{t_0}^{\tau }  d\tau' g
(\tau -\tau') v (\tau') \notag \\
& = \int_{t_{0}}^{t}d\tau' \widetilde{g} \left(\tau-\tau'\right)
\overline{u}^{\ast }\left( \tau \right) ,  \label{v-e}
\end{align}
\end{subequations}
subjected to the initial condition $u (t_0) =1$, and
$\overline{u}(\tau)\equiv u(t+t_0-\tau)$. Note that the integral
kernels in the above equations involve non-perturbatively the time
correlation functions of the reservoirs: $g(\tau-\tau')$ and $
\widetilde{g}(\tau-\tau')$. These two time-correlation functions
characterize all the non-Markovian memory structures between the
cavity and the reservoir. By defining the spectral density of the
reservoir $J(\omega) =2\pi \sum_{k}|V_{k}|^{2}\delta (\omega -\omega
_{k})$, the time-correlation functions can be expressed as
\begin{subequations}
\label{correla}
\begin{align}
g( \tau -\tau') &=\int_{0}^{\infty }\frac{d\omega}{2\pi} J\left(
\omega \right) e^{-i\omega ( \tau -\tau')} ,   \\
\widetilde{g}(\tau -\tau') &=\int_{0}^{\infty }\frac{d\omega}{2\pi} J(
\omega) \overline{n}(\omega ,T)e^{-i\omega ( \tau -\tau')} ,
\end{align}
\end{subequations}
where $\overline{n}(\omega ,T) =\frac{1}{e^{\hbar \omega /k_{B}T}
-1}$ is the average number distribution of the reservoir thermal
excitation at the initial time $t_0$. If the reservoir spectrum is
continuous, $V_{k} \rightarrow V(\omega)$, we have $J(\omega)=2\pi
g(\omega)|V(\omega)|^2$ where $g(\omega)$ is the density of state
for the reservoir.

The master equation (\ref{eme}) is exact, far beyond the BM
approximation widely used in quantum optics.  The back-reaction
effect between the system and environment is fully taken into
account by the time-dependent coefficients, $\omega'_0 (t) $,
$\kappa (t) $ and $\widetilde{\kappa}(t) $, in the master equation
(\ref{eme}) through the integrodifferential equations (\ref{uv-eq}).
In fact,  we can directly solve from Eq.~(\ref{uv-eq})
\begin{subequations}
\label{wv-eq}
\begin{align}
&\dot{u}(t)u^{-1}(t)=-i\omega_0 -\int^t_{t_0} d\tau
g(t-\tau)u(\tau) u^{-1}(t), \\
&v\left( t\right) =\int_{t_{0}}^{t}d\tau _{1}\int_{t_{0}}^{t}d\tau
_{2}\text{ }\overline{u}\left( \tau _{1}\right)\widetilde{g}\left(
\tau _{1}-\tau _{2}\right) \overline{u}^{\ast }\left( \tau
_{2}\right) . \label{v-eq}
\end{align}
\end{subequations}
Then these time-dependent coefficients in the exact master equation
can be simplified as
\begin{subequations}
\label{coefss}
\begin{align}
& \omega'_0 (t) =\omega_0 + {\rm Im}[w(t)] ,~~~
\kappa \left( t\right) ={\rm Re}[w(t)],  \label{coefss-a} \\
& \widetilde{\kappa}(t) =\dot{v}(t)+2 v(t){\rm Re}[w(t)] .
\label{coefss-b}
\end{align}
\end{subequations}
with $w(t)= \int^t_{t_0} d\tau g(t-\tau)u(\tau) u^{-1}(t)$, which
can be calculated by solving $u(\tau)$ non-perturbatively from
Eq.~(\ref{u-e}). Thus, the non-Markovian memory structure is
non-perturbatively built into the integral kernels in these
equations.  The expression of the integrodifferential equation
(\ref{uv-eq}) shows that $u(t)$ is just the propagating function of
the cavity field (the retarded Green function in nonequilibrium
Green function theory \cite{Kad62}), and $v(t)$ is the corresponding
correlation (Green) function which is also determined by $u(\tau)$
[see the equation (\ref{v-eq})]. Therefore, the exact master
equation for cavity reduced density operator depicts the full
nonequilibrium dynamics of the cavity system.

In fact, the exact master equation presented here simply covers the
exact master equation at zero-temperature we derived very recently
\cite{An07042127,An090317}. Taking the zero-temperature limit $T=0$,
then $\overline{n}\left(\omega ,T\right) =0$ so that $\widetilde{g}
(\tau -\tau{'}) =0$. As a result, we have $v (t) =0$ and therefore
the coefficient $\widetilde{\kappa}(t) =0$. The master equation is
simply reduced to
\begin{align}
\dot{\rho}\left( t\right) =&-i\omega'_0(t) \left[ a^{\dag}a,\rho(t)
\right] \notag \\ & +\kappa \left( t\right) \big\{ 2a\rho(t)
a^{\dag}-a^{\dag}a\rho(t) -\rho(t) a^{\dag}a \big\} ,  \label{emeT0}
\end{align}%
with the temperature-independent coefficients $\omega'_0(t),
\kappa(t)$ obeying the same equation (\ref{coefs-a}-\ref{coefs-b})
through Eq.~(\ref{u-e}). Eq.~(\ref{emeT0}) is the exact master
equation for cavity fields coupled to the vacuum fluctuation (i.e.
the zero temperature limit) \cite{An07042127,An090317}.

\subsection{Reduce to BM limit}

Interestingly, the exact master equation (\ref{eme}) has the same
form as the BM master equation in the literature \cite{Car93}. The
difference is the coefficients in the master equation. Here all the
coefficients are time-dependent and determined by
integrodifferential equations of motion (\ref{uv-eq}) or
(\ref{wv-eq}), which non-perturbatively takes into account the
back-reaction effects of the reservoir on the cavity field. While
the coefficients in the BM master equation are all time-independent
that ignores all the memory effects between the cavity and the
reservoir and are determined under the perturbation approximation up
to the second order of the coupling $V(\omega)$ and then taking the
Markov limit.

Explicitly, since the time correlation functions $g(\tau-\tau')$ and
$\widetilde{g}(\tau-\tau')$ are already proportional to
$|V(\omega)|^2$, taking approximately the time-dependent
coefficients in Eq.~(\ref{coefss}) up to the second-order of the
coupling $V(\omega)$ means that the propagating functions $u(\tau)$
and $u^{-1}(t)$ in the right hand side of Eq.~(\ref{wv-eq}) should
be approximated only up to the zero-order: $u_0(\tau) =
e^{-i\omega_0 (\tau-t_0)}$ and $u^{-1}_0(t) = e^{i\omega_0
(t-t_0)}$. This leads to
\begin{subequations}
\label{wv-e2}
\begin{align}
&\dot{u}(t)u^{-1}(t)\simeq -i\omega_0 - \int^t_{t_0} d\tau
\int_0^\infty \frac{d\omega}{2\pi}
J(\omega)e^{-i(\omega-\omega_0)(t-\tau)} , \label{w-e2} \\
&\dot{v}(t) \simeq  2 \int^t_{t_0} d\tau \int_0^\infty
\frac{d\omega}{2\pi} J(\omega) \bar{n}(\omega, T)
\cos[(\omega-\omega_0)(t-\tau)]. \label{vd-e}
\end{align}
\end{subequations}
The term $2v(t){\rm Re}[w(t)]$ in (\ref{coefss-b}) is proportional
to $|V(\omega)|^4$ and should be ignored in the same approximation.
Then the coefficients of Eq.~(\ref{coefs}) or (\ref{coefss}) are
reduced to
\begin{subequations}
\label{bcoefs1}
\begin{align}
&\omega'_0 (t)\simeq \omega_0 - \int^t_{t_0} d\tau \int_0^\infty
\frac{d\omega}{2\pi} J(\omega)\sin [(\omega-\omega_0)(t-\tau)], \\
&\kappa(t) \simeq  \int^t_{t_0} d\tau \int_0^\infty
\frac{d\omega}{2\pi} J(\omega)\cos [(\omega-\omega_0)(t-\tau)], \label{bcoefs1-a} \\
&\widetilde{\kappa}(t) \simeq 2 \int^t_{t_0} d\tau \int_0^\infty
\frac{d\omega}{2\pi} J(\omega) \bar{n}(\omega, T)
\cos[(\omega-\omega_0)(t-\tau)]. \label{bcoefs1-b}
\end{align}
\end{subequations}
Substituting these coefficients into Eq.~(\ref{eme}) results in the
master equation of the cavity field in the perturbation
approximation up to the second order of the coupling constant
between the cavity and the reservoir. This perturbative master
equation is a good approximation only for the dissipation and noise
dynamics of the cavity mode in the weak coupling regime.

To reproduce the standard BM master equation with the
time-independent coefficients, one needs further to take the Markov
limit where $t$ is the typical time scale for the dynamics of the
system and the $t'$ integration is dominated by much shorter time
characterizing the decay of reservoir correlations
\cite{Car93,An090317}. In other words, one can take $t'$ integration
to infinity in the equation (\ref{bcoefs1}),
\begin{align}
\lim_{t \rightarrow \infty }\int^{t-t_0}_0 d t'e^{\pm i(\omega
-\omega_0 )t'}= \pi \delta (\omega -\omega_0) \mp i\frac{\cal
P}{\omega -\omega_0} , \label{iden}
\end{align}
where ${\cal P}$ denotes the principle value of the integral. As a
result, all the coefficients in the master equation, given by
(\ref{bcoefs1}), become time-independent:
\begin{subequations}
\label{scoefs13}
\begin{align}
&\omega'_0 =\omega_0 + \delta \omega_0, ~~ \kappa = \pi
g(\omega_0)|V(\omega_0)|^2=J(\omega_0)/2 ,
\label{scoefs13-a} \\
& \widetilde{\kappa} = 2\pi g(\omega_0)|V(\omega_0)|^2 \bar{n}(\omega_0,
T)=2\kappa \bar{n}(\omega_0, T) ,\label{scoefs13-c}
\end{align}
\end{subequations}
and the frequency shift $\delta \omega_0 ={\cal P}\int_{0}^{\infty}
d\omega \frac{g(\omega)|V(\omega)|^2}{ \omega -\omega _0}$. Then the
exact master equation (\ref{eme}) is reduced to
\begin{align}
\dot{\rho}&(t)=  -\frac{i}{\hbar}(\omega_0 + \delta \omega_0)
[a^\dag a, \rho(t)] \notag \\ & + \kappa \big[ 2 a\rho (t)a^\dag  -
a^\dag a\rho (t)
- \rho(t) a^\dag a \big] \notag \\
&  + 2\kappa \bar{n}(\omega_0, T) \big[ a^{\dag }\rho(t)a + a \rho
(t)a^\dag - a^\dag a \rho (t) - \rho(t)a a^\dag  \big].
\label{smrme}
\end{align}
Eq.~(\ref{scoefs13}) with (\ref{smrme}) reproduces exactly the BM
master equation in quantum optics for a single cavity mode
interacting with a thermal field \cite{Car93}.

\section{Exact solution of the master equation}

In this section, we shall present the exact solution of master
equation for some physical observables and also for the reduced
density operator of the cavity.

\subsection{Exact solution for some physical observables}

The main physical observables for a cavity are the decays of the
mean mode amplitude and the average photon number inside the cavity.
The mean mode amplitude of the cavity field is defined by $\langle
a(t) \rangle =$tr$[ a \rho (t)] $. From the exact master equation
(\ref{eme}), it is easy to find that $\langle a(t) \rangle $ obeys
the equation of motion
\begin{align}
\langle \dot{a}(t) \rangle =-[ i\omega'_0(t) +\kappa(t)] \langle a
(t)\rangle =\frac{\dot{u}\left( t\right) }{%
u\left( t\right) }\langle a\left( t\right) \rangle .
\end{align}%
which has the exact solution:
\begin{equation}
\langle a (t) \rangle =u (t) \langle a (t_0) \rangle . \label{ma}
\end{equation}%
That is, the time evolution and the decay behavior of the mean mode
amplitude is totally determined by $u (t)$. Eq.~(\ref{ma}) clearly
indicates that $u(t)$ is the propagating function characterizing
the time evolution of the cavity field.

Another important physical observable is the average particle number
inside the cavity, which is defined by $n (t) =$tr$[ a^{\dag}a
\rho(t)]$. From the exact master equation, it is also easy to find
that
\begin{equation}
\dot{n} (t) =-2\kappa (t) n (t)
+\widetilde{\kappa} (t) . \label{diff-equ-nt}
\end{equation}%
On the other hand, Eq.~(\ref{coefs-b}) can be rewritten as
\begin{align}
\dot{v}(t) =-2\kappa(t)v (t) +\widetilde{\kappa} (t),
\end{align}%
with $-2\kappa(t)=[\dot{u}/u (t)+{\rm H.c.}]$. Combing these
equations together, we obtain the exact solution of $n(t)$ in terms
of $u(t) $ and $v(t) $:
\begin{equation}  \label{an}
n(t) =u(t)  n(t_0) u^{\ast}(t) +v(t) .
\end{equation}%
This relationship is similar to the fermion case we derived recently
\cite{Jin09101765}. In fact, the above solution is a result of the
correlated Green function in nonequilibrium Green function theory
\cite{Kad62}. Since $v(t)$ is also determined by $u(t)$ as we can
see from Eq.~(\ref{v-eq}), both the mean mode amplitude and the
average photon number inside the cavity are completely solved by the
propagating function $u(t)$.

If we take the BM limit in which all the coefficients in the master
equation are reduced to $\omega'_0=\omega_0+\delta \omega_0,
\kappa=J(\omega_0)/2, \widetilde{\kappa}=2\kappa
\overline{n}(\omega_0,T)$ with $\delta \omega_0=P\int_{0}^{\infty
}\frac{d\omega}{2\pi} \frac{J\left( \omega \right) }{\omega
-\omega_0}$ [see Eq.(\ref{scoefs13})], then the resolution of
$u\left( t\right) $ simply becomes
\begin{equation}
u_{\rm BM}\left( t\right) =e^{-(i\omega'_0+\kappa) (t-t_0)} .
\label{u-bm}
\end{equation}%
Correspondingly, the evolution of the mean mode amplitude in the BM
limit is
\begin{equation}
\langle a\left( t\right) \rangle_{\rm BM} =e^{-(i\omega'_0+\kappa)
(t-t_0)}\langle a\left( t_{0}\right) \rangle , \label{am-bm}
\end{equation}%
which shows an exponential decay in time. Substituting (\ref{u-bm})
into (\ref{v-e}), we have the BM solution of $v(t)$:
\begin{equation}
v_{\rm BM}\left( t\right) =\overline{n}\left( \omega _{0},T\right)
\big[ 1-e^{-2\kappa (t-t_0)} \big] .  \label{v-bm}
\end{equation}
Thus the average photon number in the BM limit is simply given by,
\begin{equation}
n_{\rm BM}(t) =n(t_0)e^{-2\kappa (t-t_0)} +\overline{n}\left( \omega
_{0},T\right)\big[ 1-e^{-2\kappa (t-t_0)} \big] . \label{n-bm}
\end{equation}%
These results reproduce all the BM solutions in weak coupling regime
\cite{Car93}. However, the exact solutions of Eqs.~(\ref{ma}) and
(\ref{an}) allow us to explore the non-Markovian dynamics of the
cavity systems not only in the weak coupling regime but also in the
strong coupling regime. The explicit difference between the
non-Markovian and Markov dynamics can be seen by comparing the
solution of Eqs.~(\ref{ma}) and (\ref{an}) with (\ref{am-bm}) and
(\ref{n-bm}).

\subsection{Exact Solution of the reduced density operator}

\begin{table*}[tbp]
\caption{Coefficients in the exact solution of the propagating
function}
\begin{center}
\begin{tabular}{|c|c|c|c|c|}
\hline & $A(t) $ & $ B(t) $ & $C(t) $ & $D(t)$\\ \hline Exact result
& $\frac{1}{1+v(t)}$ & $\frac{u(t)}{1+v(t)}$ & $\frac{v(t)}{1+v(t)}$
& $1-\frac{|u(t)|^2}{1+v(t)}$ \\ \hline $T \rightarrow 0$ limit & $1
$ & $u(t) $ & $0$ & $1-|u(t)|^2$
\\ \hline BM limit$^*$ & $
\frac{1}{1+v_{\rm BM}(t) }$ & $\frac{u_{\rm BM}(t)}{1+ v_{\rm BM}(t)
}$ & $\frac{v_{\rm BM}(t)}{1+
v_{\rm BM}(t) }$ & $ 1 - \frac{|u_{\rm BM}(t)|^2}{1+v_{\rm BM}(t)}$ \\
\hline
\end{tabular} \\ \ \\
$^*$where $u_{\rm BM}(t)$ and $v_{\rm BM}(t)$ are given by
Eqs.~(\ref{u-bm}) and (\ref{v-bm}).
\end{center}\label{table}
\end{table*}

In fact, the reduced density operator can be also explicitly obtained
through the coherent state representation. The reduced density
matrix in the coherent state representation is given by
\begin{equation}
\rho (t) =\int d\mu ( \alpha_{f} ) d\mu ( \alpha'_{f}) \rho (
\alpha^*_{f},\alpha'_{f},t) | \alpha_{f}\rangle \langle \alpha'_{f}|
, \label{srdm}
\end{equation}%
where $ d\mu (\alpha)\equiv \frac{d\alpha^*d\alpha}{2\pi
i}e^{-|\alpha|^2}$ is the integrate measure in the complex space of
the coherent state $|\alpha \rangle=e^{\alpha a^\dag}|0\rangle$
\cite{Zhang9062867}, and $\rho (\alpha^*_{f},\alpha'_{f},t)=\langle
\alpha_{f}|\rho (t)| \alpha'_{f}\rangle $ can be obtained from the
exact master equation:
\begin{align}
\rho (\alpha^*_{f},\alpha'_{f},t)= & A(t)\int d\mu ( \alpha_{i})
d\mu ( \alpha'_{i}) \rho ( \alpha^*_{i},\alpha'_{i},t_{0}) \notag
\\ & \times \exp\big\{\alpha^*_{f}B(t) \alpha_{i}+\alpha^*_{f}C(t)
\alpha'_{f}\notag \\
& ~~~~~~~~~~ +\alpha^*_{i}D(t) \alpha_{i} +{\alpha'}^*_{i}B^*(t)
\alpha'_{f} \big\}, \label{srdmc}
\end{align}%
where $\rho (\alpha^*_{i},\alpha'_{i},t_{0})=  \langle
\alpha_{i}|\rho ( t_{0}) | \alpha'_{i}\rangle $ is the initial state
$\rho(t_0)$ in the coherent state representation. The exponential
function in the integral is the propagating function of the reduced
density operator, which fully determines the time evolution of the
reduced density matrix in the coherent state representation. The
time-dependent coefficients $A(t), B(t), C(t)$ and $D(t)$ in the
propagating function are listed in table \ref{table} for various
cases. Thus, for a given initial state $\rho(t_0)$, the
corresponding analytical solution of the reduced density operator
for an arbitrary coupling to the reservoir at an arbitrary
temperature can be obtained from Eqs.~(\ref{srdm}) and
(\ref{srdmc}). Here we shall consider a few typical initial states:
the vacuum state, the coherent state and the mixed state.

\subsubsection{Initial vacuum state}

First, we consider the cavity initially in the vacuum state,
\begin{equation}
\rho \left( t_{0}\right) =\left\vert 0\rangle \langle 0\right\vert .
\end{equation}%
From Eqs.~(\ref{srdm}) and (\ref{srdmc}), it is not difficult to
obtain the exact reduced density operator at arbitrary time $t$,
\begin{align}
\rho (t) &=A (t) \sum_{n=0}^{\infty }[C(t)]^{n} |n\rangle \langle n
| \notag \\
&=\sum_{n=0}^{\infty }\frac{[ v(t)]^{n}}{[ 1+v(t)]^{n+1}}| n\rangle
\langle n| .  \label{v-ms}
\end{align}%
This is a completely mixed state in terms of the Fock space of the
cavity, with the average photon number
\begin{align} n(t) =v(t) .
\end{align}
That is, there is no existence of a coherent field in the cavity if
it is initially empty and also no external driving field is applied
to the cavity. It indicates that the coupling to the reservoir makes
the cavity continuously but randomly gain the energy (photons) from
the reservoir, and eventually reach to a steady mixed state with a
constant average photon number $n(t)=v(t \rightarrow \infty)$.
However, we must point out that this steady limit is generally
different from the BM limit since $v(t \rightarrow \infty) \neq
\overline{n}(\omega_0, T)$. The difference is a manifestation of the
non-Markovian effect that we shall demonstrate numerically in the
next section.

At zero-temperature limit, $v(t)=0$, then
\begin{equation}
\rho \left( t\right)|_{T=0}=\left\vert 0\rangle \langle 0\right\vert
=\rho \left( t_0 \right) .
\end{equation}%
 In other words, the vacuum state remains unchanged if the reservoir
temperature is zero. This is a trivial physical consequence since at
zero temperature, the reservoir is also in the vacuum state. Then no
photon can be exchanged between the cavity and the reservoir so that
the cavity remains unchanged. On the other hand, in the BM limit, we
have
\begin{align}
\rho_{\rm BM} ( t )&= \sum_{n}\frac{[v_{\rm BM}(t)]^{n}}{[ 1+v_{\rm
BM}(t)]^{n+1}} | n\rangle \langle n|  \notag \\
&\overset{t\rightarrow \infty}{=} \sum_{n}\frac{[\overline{n}(
\omega_0,T)] ^{n}}{[ 1+\overline{n}( \omega_0,T)]^{n+1}} | n\rangle
\langle n |,  \label{s1bm}
\end{align}
which is a thermal equilibrium state with the average photon number
$ n_{BM}(t) \rightarrow \overline{n}( \omega _0,T) $ at temperature
$T$. It is expected that the solution of Eq.~(\ref{v-ms}) can be
very different from that of Eq.~(\ref{s1bm}) in the strong coupling
regime.

\subsubsection{Initial coherent state}

Next, we consider an initial coherent state,
\begin{equation}
\rho \left( t_{0}\right) =e^{-|\alpha _{0}| ^{2}} \left\vert \alpha
_{0}\rangle \langle \alpha _{0}\right\vert ,
\end{equation}%
From Eqs.~(\ref{srdm}) and (\ref{srdmc}), we find that the reduced
density operator at time $t$ becomes
\begin{align}
\rho(t) = & \exp\Big\{\frac{|\alpha(t)|^2}{1+v(t)}\Big\}
\sum_{n=0}^{\infty } \frac{[ v(t)]^{n}}{[1+v(t)]^{n+1}} \notag \\
&~~~~~~~~~~~ \times \Big|\frac{\alpha(t)}{1+v(t)},n \Big\rangle
\Big\langle n, \frac{\alpha(t)}{1+v(t)}\Big| , \label{mcs}
\end{align}%
where $|\frac{\alpha(t)}{1+v(t)}, n\rangle \equiv
\exp[\frac{\alpha(t)} {1+v(t)}a^\dag]|n\rangle$ is defined as a
generalized coherent state, and $\alpha(t)=u(t)\alpha_0$. It is
interest to see that Eq.~(\ref{mcs}) is indeed  a mixed state of
generalized coherent states $|\frac{\alpha(t)}{1+v(t)}, n \rangle$.
The average photon number in this state is given by
\begin{align}
n(t)=|\alpha(t)|^2 + v(t) ,
\end{align}
as we expected.

In the weak coupling regime, $u(t)$ generally decays to zero. The
steady state limit of the above state will be the same as that in
the case 1, and the asymptotic average photon number is also given
by $n(t)=v(t \rightarrow \infty)$. The corresponding reduced density
operator asymptotically becomes a completely mixed state of
Eq.~(\ref{v-ms}). This solution shows an exact decoherence process
in the cavity. The decoherence arises from two sources, the mean
mode amplitude damping characterized by the decay behavior in terms
of the propagating function through the solution
$\alpha(t)=u(t)\alpha_0$, and the thermal-fluctuation-induced noise
effect characterized by $v(t)$, as we can see from Eq.~(\ref{mcs}).
The later describes a process of randomly loss or gain thermal
energy from the reservoir, up to the initial temperature of the
reservoir.

However, in the strong coupling regime, $u(t)$ may not decay to
zero, as we shall show explicitly in the numerical calculation in
the next section. Then the reduced density operator remains as a
mixed coherent state. On the other hand, at zero-temperature limit
($v(t)=0$), the reduced density operator at time $t$ is given by
\begin{align}
\rho \left( t\right) |_{T=0} = e^{-|\alpha (t) |^{2}} |\alpha (t)
\rangle \langle \alpha (t)|.
\end{align}%
In other words, the cavity can remain in a coherent state in the
zero temperature limit. These two features [$u(t)$ may not decay to
zero in the strong coupling regime and $v(t)=0$ at $T=0$] indicate
that enhancing the coupling between the cavity and the reservoir and
meantime lowing the initial temperature of the reservoir can
significantly reduce the decoherence effect in cavity system.

On the other hand, in BM limit the reduced density operator is
always reduced to a thermal state after a long time:
\begin{equation}
\rho_{\rm BM} (t\rightarrow \infty) =\sum_{n=0}^{\infty }\frac{[
\overline{n}( \omega
_{0},T) ] ^{n}}{[ 1+\overline{n}( \omega _0,T ) %
] ^{n+1}}| n\rangle \langle n |,  \label{rdm-bm}
\end{equation}
This is because $u_{\rm BM}(t\rightarrow \infty)$ must approach to
zero, see Eq.~(\ref{u-bm}). It shows that the results in BM limit
can be qualitatively different from the exact solution. These
analytical results reveal the underlying mechanism of
reservoir-induced decoherence in cavity dynamics, which may provide
some insight how to control the decoherence dynamics in such
systems.

\subsubsection{Initial mixed state}

The last special case we shall consider is the cavity in an
initially mixed state,
\begin{equation}
\rho \left( t_{0}\right) =\sum_{n=0}^{\infty }\frac{[n(t_0)]^{n}}
{[1+n(t_0)]^{n+1}} | n\rangle \langle n|,
\end{equation}%
with the average initial photon number $n(t_0)$. Using the solution
of the master equation, we have
\begin{align}
\rho (t) =&\frac{A(t) }{1+n(t_0)-n(t_0)D(t)}  \notag \\
&\times \sum_{n=0}^{\infty }\left( \frac{n(t_0)|B(t)|^{2}}{1+n(t_0)-
n(t_0)D(t)}+C(t) \right)^{n}\left\vert n\rangle \langle n\right\vert
\notag \\
=&\sum_{n=0}^{\infty }\frac{\left[ n\left( t\right)  \right]
^{n}}{\left[ 1+ n\left( t\right) \right] ^{n+1}}\left\vert n\rangle
\langle n\right\vert ,
\end{align}%
where \begin{align}  n(t) =|u\left( t\right)|^2 n(t_0) +v\left(
t\right)
\end{align} is the average photon number of the cavity during time
evolution. It shows that if the cavity is initially in a mixed
state, it will remain in a mixed state, with different particle
number distribution varying in time. In the weak coupling regime,
$u(t)$ approaches to zero so that the reduced density matrix at
steady limit will reach to the same state as in the other two cases.
In the strong coupling regime, $u(t)$ may not decay to zero so that
the average photon number in cavity $n(t)$ can be very different
from the weak coupling regime, namely very different from the BM
limit, even though both the exact reduced density matrix and its BM
limit are mixed states.

In the next section, we will numerically demonstrate these dynamics
for some specifically given spectral density $J(\omega)$.

\section{Numerical analysis of the exact non-Markovian dynamics}

\begin{figure*}
\includegraphics[width=16cm]{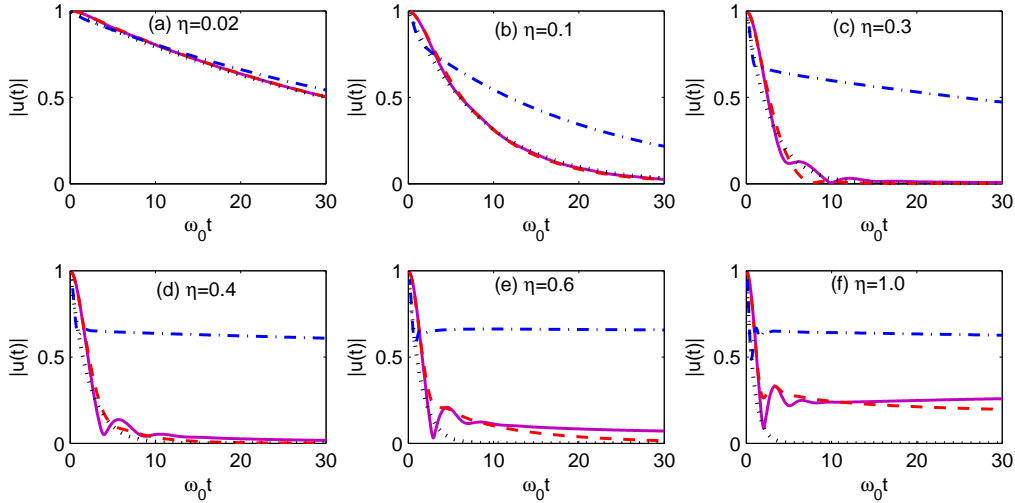}\newline
\caption{Comparison of the exact solution of $|u(t)|$ for sub-Ohmic
(solid Magenta line), Ohmic (dashed red line) and super-Ohmic
(dash-dotted blue line) with the corresponding BM limit (dotted
black line) in the weak coupling ($\eta \leq 0.3)$ and the strong
coupling ($\eta \geq 0.4)$ regimes. Here we have taken the
parameters $\omega_0=21.5$GHZ, $\omega_c=\omega_0$.} \label{ut}
\end{figure*}

To explicitly see the non-Markovian memory effect in the cavity
system when it strongly couples with a reservoir, we consider a
general spectral density of the bosonic environment
\begin{equation}
J\left( \omega \right) =2\pi\eta \omega \left( \frac{\omega }{\omega _{c}}%
\right) ^{s-1}e^{-\omega /\omega_{c}},  \label{sd}
\end{equation}%
where $\eta $ is the dimensionless coupling strength between the system and
reservoir, $\omega_{c}$ is the cutoff frequency of the spectrum. The
parameter $s$ classifies the environment as sub-Ohmic ($0<s<1$), Ohmic ($s=1$%
), and super-Ohmic ($s>1$). In the following, we set the value of $s$ to $%
1/2 $, $1$, and $3$ respectively.

\begin{figure*}
\includegraphics[width=16cm]{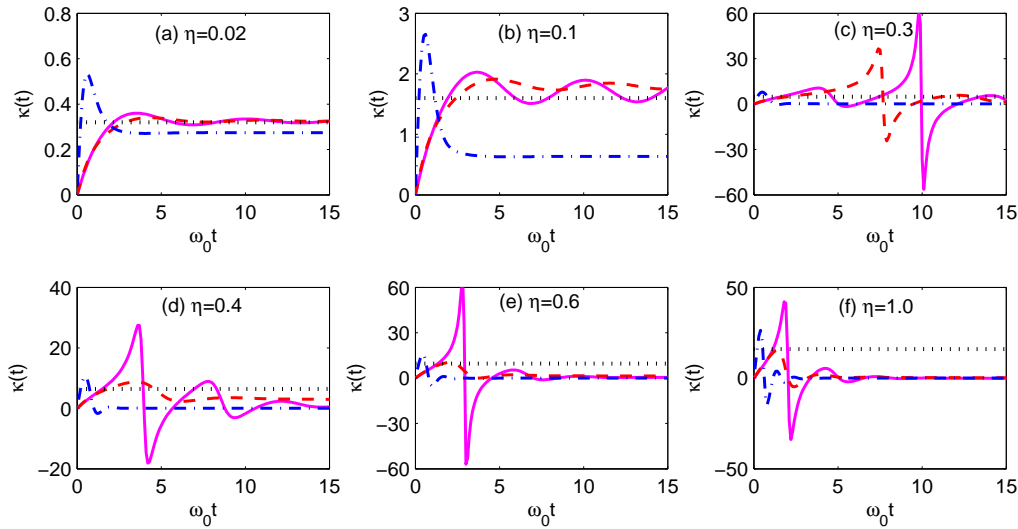}\newline
\caption{Comparison of the exact solution of $\kappa(t)$ for
sub-Ohmic (solid Magenta line), Ohmic (dashed red line) and
super-Ohmic (dash-dotted blue line) with the corresponding BM limit
(dotted black line) in the weak coupling ($\eta \leq 0.3)$ and the
strong coupling ($\eta \geq 0.4)$ regimes. Here we have taken the
parameters $\omega_0=21.5$GHZ, $\omega_c=\omega_0$.} \label{kt}
\end{figure*}

\begin{figure*}[tbp]
\includegraphics[width=16cm]{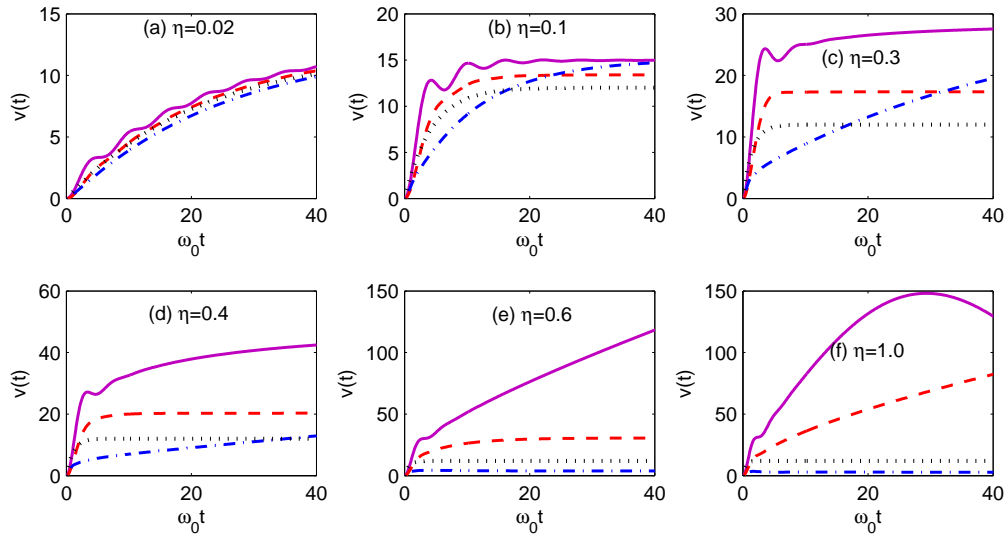}\newline
\caption{Exact solution of $v(t)$ for sub-Ohmic (solid Magenta
line), Ohmic (dashed red line) and super-Ohmic (dash-dotted blue
line) with the corresponding BM limit (dotted black line) in the
weak coupling ($\eta \leq 0.3)$ and the strong coupling ($\eta \geq
0.4)$ regimes. Here $\omega_0=21.5$GHZ, $\omega_c=\omega_0$, and
$T=2$K.} \label{vt}
\end{figure*}

\begin{figure*}[tbp]
\includegraphics[width=11cm]{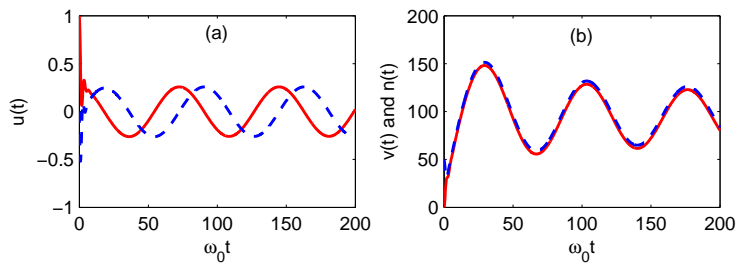}\newline
\caption{Long time behavior of $u(t)$, $v(t)$ and $n(t)$ for
sub-Ohmic spectral density with $\eta=0.1$. Figure (a) shows ${\rm
Re}[u(t)]$ (red solid line) and ${\rm Im}[u(t)]$ (dashed blue line).
Figure (b) shows $v(t)$ (red solid line) and $n(t)$ (dashed blue
line). Here $\omega_0=21.5$GHZ, $\omega_c=\omega_0$, $T=2$K and
$n(t_0)=50$.} \label{subohmic-eta=1}
\end{figure*}

\begin{figure*}[tbp]
\includegraphics[width=11.5cm]{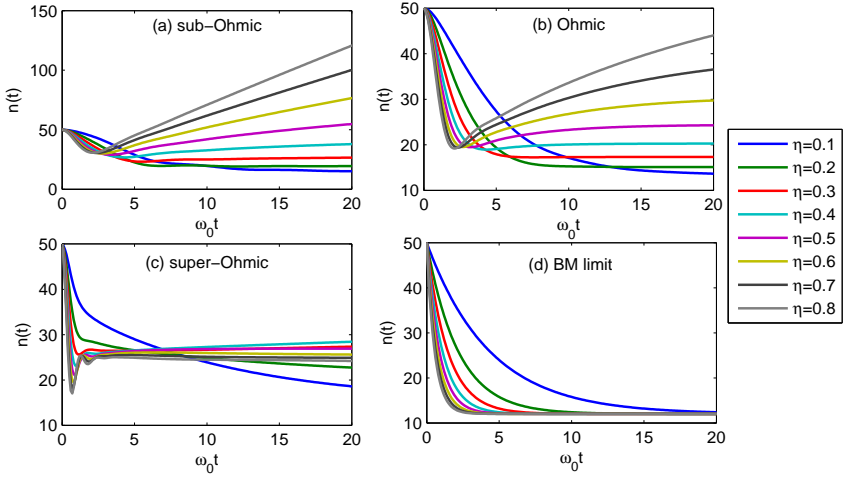}\newline
\caption{Time evolution of the average particle number $n(t)$. Here
$\omega_0=21.5$GHZ, $\omega_c=\omega_0$, $T=2$K, and $n(t_0)=50$.}
\label{nt}
\end{figure*}

\begin{figure*}[tbp]
\includegraphics[width=11.5cm]{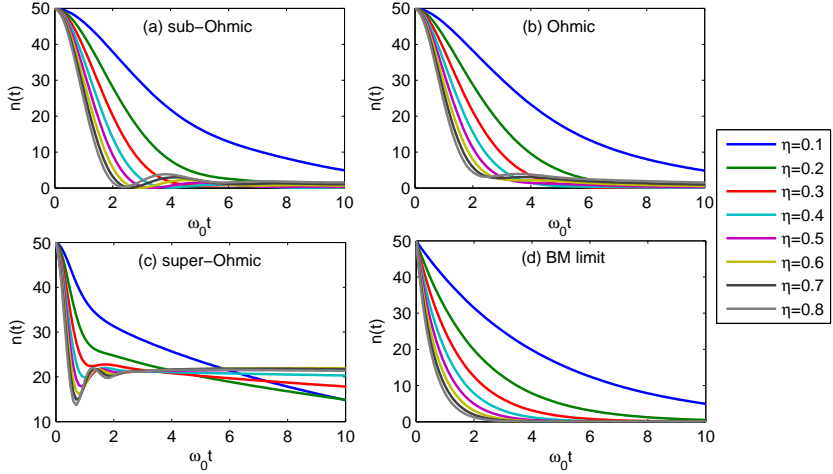}\newline
\caption{Time evolution of the average particle number $n(t)$. Here
$\omega_0=21.5$GHZ, $\omega_c=\omega_0$, $T=0.002$K, and
$n(t_0)=50$.} \label{ntlT}
\end{figure*}

For a single-mode cavity with the frequency in the optical regime,
it is well known that the BM approximation works very well for a
usual thermal environment. However, for a structured reservoir, such
as the CROW, the coupling strength between the cavity and reservoir
can be controlled by changing the geometrical parameters of the
defect cavity and the distance between the cavity and the CROW
\cite{cavity with CROW -7}. Then the BM approximation or the
perturbative approximation must be reexamined. Furthermore, when the
cavity frequency lies in the microwave regime, the temperature of
the environment also becomes non-negligible. Specifically we take
the cavity frequency to be $\omega _{0}=21.5$ GHz, i.e., $\hbar
\omega_{0}=13.83{\mu}$eV, and the temperature is taken at $T=2$K so
that $k_{B}T=172.3{\mu}$eV $\approx 12.5 \hbar\omega_{0}$
\cite{experimental parameters}. With these experimental parameters
as input, we numerically calculate the exact dissipative dynamics of
the cavity for three different spectral densities with different
coupling strength $\eta$ in Eq.~(\ref{sd}). In addition, the cutoff
frequency $\omega_{c}$ in Eq.~(\ref{sd}) is taken roughly the same
order as the cavity frequency $\omega_{0}$, i.e. $\omega_{c}\approx
\omega_{0}$. The detailed numerical results are plotted in
Figs.~\ref{ut}-\ref{ntlT}. Note that the BM limits are the same for
three different spectral densities when we take the cut-off
frequency $\omega_c=\omega_0$ since $u_{\rm BM}(t)$, $v_{\rm BM}(t)$
and $n_{\rm BM}(t)$ only depends on $\kappa
=J(\omega_0)/2=\pi\eta\omega_0{e^{-1}}$ for all three different
spectral densities [see Eq.~(\ref{sd})]. From Eqs. (\ref{u-bm}),
(\ref{v-bm}) and (\ref{n-bm}), we can analytically know that $u(t)$,
$v(t)$ and $n(t)$ in the BM limit are monotonous change in time as
the increases of the coupling $\eta$. However, in the exact cases,
we will show that the results have qualitative changes from the weak
to strong coupling regimes.

In Fig.~\ref{ut}, we show the exact solutions for the absolute value
of $u(t)$ (i.e. the amplitude of the propagating function $u(t)$
which characterizes the time evolution of the amplitude of the
cavity field through the relation $\langle a(t)\rangle=u(t)\langle
a(t_0)\rangle$) for Ohmic, sub-Ohmic and super-Ohmic spectral
densities in the weak and strong coupling cases with a comparison to
the BM limit. For a given coupling strength $\eta$, comparing the
behavior of the exact $u(t)$ of different spectral densities with
its BM limit, we see that in the very weak coupling limit ($\eta
=0.02$), the difference between the exact amplitude of $u(t)$ and
the BM limit is very small for all the three spectral densities.
With the increasing of $\eta$, the difference becomes more and more
visible. The large difference between the exact result and the BM
limit for a strong coupling is a significant manifestation of the
non-Markovian memory effect.

For a given spectral density, comparing the behavior of $|u(t)|$
among different coupling strengths, we find that roughly for $\eta <
0.3$, $|u(t)|$ decays almost monotonously for all the three spectral
densities, except for a short-time oscillation in the beginning. In
general, a weak coupling to the reservoir always induces an
amplitude damping to the cavity field. However, when $\eta > 0.3$,
besides a short-time oscillation and decay, $|u(t)|$ may approach to
a nonzero stationary value. In other words, in the strong coupling
regime, the non-Markovian memory effect can result in a long-time
qualitatively change to the time evolution of the cavity field,
namely it changes the cavity field from a pure damping process in
the weak coupling regime to a dissipationless process in the strong
coupling regime. In particular, this qualitative change in the exact
solution of $u(t)$ becomes the most significant for the super-Ohmic
case, and then for the sub-Ohmic and the Ohmic cases, in comparison
with the BM limit, as shown in Fig.~\ref{ut}.

In order to see cleanlier the short- and long-time behaviors of
$u(t)$, we plot $\kappa(t)=-{\rm Re} [\dot{u}(t)/u(t)]$ in
Fig.~\ref{kt} which is the decay coefficient in the master equation
(\ref{eme}) that describes the energy dissipation of the cavity. The
short-time rapidly increasing of $\kappa(t)$ in the beginning
indicates a fast decay of the cavity field, in agreement with the
results shown in Fig.~\ref{ut}. However, with increasing the
coupling $\eta$, in particular for $\eta \geq 0.3$, we see that the
decay coefficient $\kappa(t)$ shows a very different
time-dependence. It has a large oscillation between an equal
positive and negative bounded value and then approaches to zero. The
oscillation indicate that the cavity dissipates energy into the
reservoir first and gains the energy back from the reservoir. Then
the zero steady value indicates that after the oscillation, the
cavity becomes dissipationless. This is why $|u(t)|$ has a nonzero
steady value, as we seen in Fig.~\ref{ut}. $|u(t)|$ of the
super-Ohmic reservoir maintains a largest nonzero steady value in
the strong coupling regime because its short-time oscillation occurs
in the shortest time within which only the smallest energy is
dissipated. While, in the BM limit, $\kappa$ keeps in a nonzero
constant, namely the cavity is always in a dissipation state.

Therefore, we can conclude that the super-Ohmic reservoir contains
the strongest non-Markovian memory effect shown in the long-time
dissipationless process for the cavity field, while the sub-Ohmic
reservoir involves the strongest short-time oscillating
non-Markovian dynamics. In the previous investigations on
non-Markovian effect, one has observed the short-time oscillation
behavior in the dissipation processes in the weak coupling regime
\cite{trapped ion in engineered reservior -4}. The long-time
dissipationless stationary behavior in the strong coupling regime
may only be revealed from a nonperturbation theory, as we shown
here.

To see clearly the significance of the temperature effect of the
non-Markovian memory dynamics, we show in Fig.~\ref{vt} the time
evolution of the correlation function $v(t)$ in the weak and strong
coupling regimes and compare the results with the BM solution again.
The correlation function $v(t)$ characterizes the
temperature-dependent noise effect (the average photon correlation
through the reservoir).  Meanwhile, $v(t)$ is also the main
contribution to the average photon number inside the cavity induced
mainly by thermal noise dynamics, see Eq.~(\ref{an}).
It is shown in Table \ref{table} that, in
the zero-temperature limit $T\rightarrow{0}$, $v(t)=0$. However, at
a finite temperature, it will behave very differently for different
couplings.

In the very weak coupling limit ($\eta = 0.02$), the exact $v(t)$ is
almost the same as the BM solution, except for the sub-Ohmic
spectral density in which $v(t)$ shows a long-time oscillation, as a
weak non-Markovian memory effect. With the increasing of $\eta$, the
quantitative difference between the exact solution and its BM limit
is enlarged for all the three spectral densities. Comparing the
behavior of $v(t)$ among different $\eta$ for a given spectral
density, we find that, similar to the behavior of $|u(t)|$, a
short-time oscillation (almost invisible in Fig.~\ref{vt}) of $v(t)$
exists for all the three spectral densities. In the long-time limit,
for different coupling strengths, the stationary values of the exact
$v(t)$ for all three different spectral densities can be very
different from the BM result. The numerical result in Fig.~\ref{vt}
clearly shows that $v(t\rightarrow \infty)$ is far away from the BM
limit $v_{\rm BM}(t\rightarrow \infty)=\overline n(\omega_0, T)$ in
the strong coupling regime, as we discussed analytically in the last
section.

In particular, continuously increasing the coupling strength $\eta$
shows that the sub-Ohmic case manifests a significantly qualitative
difference from the BM limit. The time oscillation behavior of
$v(t)$ for the sub-Ohmic case is very strong and maintains for a
longer time when the coupling $\eta$ becomes larger. This long-time
oscillation behavior comes actually from the long-time periodical
oscillation of $u(t)$, as we shown in Fig. \ref{subohmic-eta=1}.
However, in the strong coupling regime, $v(t)$ of the super-Ohmic
spectral density increases slower than the BM limit with increasing
$\eta$ but the overall time evolution is qualitatively the same as
in the BM limit. In other words, the temperature-induced noise
effect becomes very important for sub-Ohmic case and then for the
Ohmic case while it may be minor for the super-Ohmic reservoir in
the strong coupling regime, which is quite different in comparison
with the solution in the weak coupling cases, see Fig.~\ref{vt}.

To show the totally non-Markovian memory effect distributing in the
amplitude damping and the noise dynamics, we examine the average
photon number inside the cavity which is analytically given by
Eq.~(\ref{an}).  From equation (\ref{an}), we see that the average
photon number contains two terms, the first term is the decay of the
initial average photon number determined by the propagating function
$u(t)$, and the second term is just the correlation function $v(t)$
as a noise effect which sensitively depends on the initial
temperature of the reservoir. The numerical solutions based on the
exact solution (\ref{an}) for three different spectral densities are
plotted in Fig.~\ref{nt} with a comparison to the BM result.

In the weak coupling regime (roughly $\eta < 0.3 )$, as we see the
exact solution of $n(t)$ shows a monotonic decay, similar to that in
the BM limit. However, in the strong coupling regime (roughly $\eta
> 0.3)$, a revival process occurs in the exact $n(t)$, namely the
average photon number decays faster in the very beginning and then
after a short-time oscillation (or no oscillation), it revives till
it reaches to a steady value. This behavior does not show up in the
BM approximation. In fact, the plots in Fig.~\ref{nt} show that the
time evolution of the average photon number undergoes a critical
transition from the weak to strong coupling regimes. The transition
regime here corresponds roughly to $\eta \simeq 0.3 $. This critical
transition results in a competition between the non-Markovian memory
effect induced dissipationless phenomena with the thermal noise
dynamics in the strong coupling in open systems.

Furthermore, this critical transition can still be seen when we
reduce the initial temperature of the reservoir to a very low value,
see Fig.~\ref{ntlT} where $T=0.002K$. The critical transition is
clearly shown up for the super-Ohmic case. For the Ohmic and
sub-Ohmic cases, the critical phenomena is not so significant but
the average photon number can reach to a very small but nonzero
steady value in the strong coupling regime. This result indicates
that the initial temperature of the reservoir may serve as a
sensitive control parameter to control the coherence photon number
in cavity dynamics in the very low temperature regime, as we have
also discussed in the analytical solution presented in the last
section.

Put all these analysis together, we find that the non-Markovian
memory effect can qualitatively change the dissipation dynamics of
the cavity field in the strong coupling regime, in particular for
the super-Ohmic reservoir. Meantime, the non-Markovian memory
effects play an significant role to the thermal noise dynamics, in
particular for the sub-Ohmic reservoir. These interesting phenomena
worth further investigation in other open systems.

\section{Conclusion}
In summary, we have solved analytically the exact master equation,
and obtained the general expression for the reduced density operator
of the cavity as well as the general solution of the mean model
amplitude and the average photon numbers in the cavity. We take
three different initial states: the vacuum state, the coherent state
and the mixed state, to show the different non-Markovian time
evolution of the cavity state. We find that (i) The solution of the
exact master equation is very different from the BM approximation
due to the non-Markovian memory effect. In the exact non-Markovian
case, different initial states may result in different evolution
states and different steady states. While the BM solution at steady
state limit are the same, independent of the initial states. (ii)
For the exact non-Markovian evolution process, temperature effect
can play an important role, and it drastically changes the thermal
noise dynamics. (iii) The decoherence of cavity coherent field
arises from both the reservoir-induced damping effect (energy
dissipation) and the temperature-dependent noise effect. Both of
them can be controlled by varying the coupling between the cavity
and the reservoir and lowing the temperature of the reservoir, as
shown in our exact solution. These exact analytical cavity dynamics
show the extreme importance of the non-Markovian effect and are
numerically demonstrated. Moreover, the numerical results show that,
the non-Markovian memory effect qualitatively changes the amplitude
damping behavior of the cavity field as well as the thermal noise
dynamics from the weak coupling regime to strong coupling regime,
which does not occur in the BM limit. In particular, we show that
the non-Markovian memory effect leads to a long-time dissipationless
process to the cavity field in the strong coupling regime, and
meantime it results in a critical transition dynamics when we vary
the cavity system from the weak to strong couplings with the
reservoir. Further investigations of all these phenomena in other
open systems are in progress.

\textit{Acknowledgements:} We thank Chan U Lei for some numerical
checks. This work is supported by the National Science Council of
ROC under Contract No. NSC-96-2112-M-006-011-MY3, NSFC with grant
No.10874151, 10935010, NFRPC with grant No. 2006CB921205; Program
for New Century Excellent Talents in University (NCET), and Science
Foundation of Chinese University.

\end{document}